\documentclass[preprint,preprintnumbers,amsmath,amssymb,floatfix,nofootinbib,superscriptaddress]{revtex4}
\usepackage[dvips]{graphicx}
\usepackage{url}
\usepackage{subfigure}
\usepackage{color}

\def\be{\begin{equation}}
\def\ee{\end{equation}}
\def\ba{\begin{eqnarray}}
\def\ea{\end{eqnarray}}

\def\mur{\mu_\mathrm{R}}
\def\muf{\mu_\mathrm{F}}

\def\POWHEGBOX{{\tt POWHEG BOX}}
\def\POWHEG{{\tt POWHEG}}
\def\SHERPA{{\tt SHERPA}}
\def\PowHel{{\tt PowHel}}
\def\aMCNLO{{\tt aMC@NLO}}
\def\MadaMCNLO{{\tt MadGraph5\_aMC@NLO}}
\def\PYTHIA{{\tt PYTHIA}}
\def\PYTHIAsix{{\tt PYTHIA6}}
\def\PYTHIAeight{{\tt PYTHIA8}}
\def\HERWIG{{\tt HERWIG}}
\def\tth{t\bar t H}
\def\mr{\mathrm}
\begin{document}          
%
%\preprint{hep-ph/yymmnnn}
%
\title{Higgs boson production in association with top quarks in the
  \POWHEGBOX{}}
\author{H.~B.~Hartanto}
\email{hartanto@physik.rwth-aachen.de}
\affiliation{Insitut f\"{u}r Theoretische Teilchenphysik und Kosmologie,
RWTH Aachen University, D-52056 Aachen, Germany}
\author{B.~J\"ager}
\email{barbara.jaeger@itp.uni-tuebingen.de}
\affiliation{Institute for Theoretical Physics, T\"ubingen University,
 Auf der Morgenstelle 14, 72076 T\"ubingen, Germany}
\author{L.~Reina}
\email{reina@hep.fsu.edu}
\affiliation{Physics Department, Florida State University,
Tallahassee, FL 32306-4350, U.S.A.}
\author{D.~Wackeroth}
\email{dow@ubpheno.physics.buffalo.edu}
\affiliation{Department of Physics, SUNY at Buffalo, Buffalo, NY 14260-1500, U.S.A.}

\date{\today}

\begin{abstract}
  We present results from the analytic calculation of $t\bar{t}H$
  hadronic production at Next-to-Leading Order in QCD interfaced with
  parton-shower Monte Carlo event generators in the \POWHEGBOX{}
  framework. We consider kinematic distributions of the top quark and
  Higgs boson at the 8 TeV Large Hadron Collider and study the
  theoretical uncertainties due to specific choices of
  renormalization/factorization scales and parton-showering
  algorithms, namely \PYTHIA{} and \HERWIG.  The importance of
  spin-correlations in the production and decay stages of a
  top/antitop quark is discussed on the example of kinematic
  distributions of leptons originating from the top/antitop
  decays. The corresponding code is now part of the public release of
  the \POWHEGBOX{}.
\end{abstract}
%\pacs{}
%\keywords{}
%
\maketitle

\section{Introduction}
\label{sec:introduction}

The associated production of a Higgs boson with a pair of top ($t$) and
antitop quarks ($\bar t$) has been receiving increased attention after
the discovery of a Higgs boson ($H$) at the CERN Large Hadron
Colllider (LHC)~\cite{Aad:2012tfa,Chatrchyan:2012ufa}. Although rare
in nature, this production mode gives direct access to the Higgs-boson
coupling to the heaviest elementary fermion, the top quark, and
provides important constraints on extensions of the Standard Model
(SM) when combined with the indirect determination of the same
coupling in Higgs-boson production via loop-induced gluon-gluon
fusion. Experimental studies of Higgs properties have already provided
strong constraints in Run
1~\cite{Chatrchyan:2012jja,Aad:2013xqa,ATLAS:2013sla}, and will be one
of the main focuses of Run 2 of the LHC.

Recent experimental analyses have explored the possibility of
detecting $t\bar{t}H$ production when the Higgs boson decays via
$H\rightarrow
b\bar{b}$~\cite{ATLAS:2012cpa,CMS:2012qaa,Chatrchyan:2013yea,ATLAS:CONF-2014-011},
$H\rightarrow\gamma\gamma$~\cite{ATLAS:CONF-2013-080,CMS:2013fda,ATLAS:CONF-2014-043},
and $H\rightarrow ZZ^*,WW^*$~\cite{CMS:2013tfa}, or in a combination
of all the previous modes~\cite{Khachatryan:2014qaa}.  Apart from
overcoming challenging background difficulties, it is important first
of all to have the signal fully under control, including the decay of
the final-state particles. With this respect, the theoretical
systematic error has to be carefully investigated and reduced, if
possible, by adding both QCD and electroweak higher-order corrections
consistently interfaced with existing parton-shower (PS) Monte Carlo
codes, e.g.  \PYTHIA{}~\cite{Sjostrand:2006za,Sjostrand:2007gs},
\HERWIG{}~\cite{Marchesini:1991ch,Corcella:2000bw} or
\SHERPA{}~\cite{Gleisberg:2008ta}, to match the complexity of real
events.

The parton-level ${\cal O}(\alpha_s^3)$ or next-to-leading order (NLO)
QCD cross section for $t\bar{t}H$ production has been first calculated
in
Refs.~\cite{Beenakker:2001rj,Beenakker:2002nc,Reina:2001sf,Reina:2001bc,Dawson:2002tg,Dawson:2003zu}
and subsequently confirmed by various collaborations using
one-loop computational tools such as
\aMCNLO~\cite{Frederix:2011zi,Hirschi:2011pa} and
\PowHel~\cite{Garzelli:2011vp,Bevilacqua:2011xh}. More recently the
${\cal O}(\alpha_s\alpha^2)$ contribution to the parton level cross
section has also been calculated including~\cite{Yu:2014cka} or
omitting~\cite{Frixione:2014qaa} QED corrections.  Effects of parton
shower and hadronization on the NLO-QCD parton-level cross sections
have been studied both in the \aMCNLO{}
framework~\cite{Frederix:2011zi} and in a private implementation of
the \POWHEG{} method by the \PowHel{}
collaboration~\cite{Garzelli:2011vp}. The two approaches have been
compared and found in good agreement in a study performed within the
LHC Higgs Cross Section Working Group~\cite{Dittmaier:2012vm}. More
recently, the original NLO-QCD calculation of
Refs. ~\cite{Reina:2001sf,Reina:2001bc,Dawson:2002tg,Dawson:2003zu} has
been interfaced with \SHERPA~\cite{Gleisberg:2008ta} and has become
part of the {\tt SHERPA-2.0.0} release.  We have now interfaced the
same routines with the
\POWHEGBOX{}~\cite{Nason:2004rx,Frixione:2007vw,Alioli:2010xd}
including the decays of the top/antitop quarks and the Higgs boson.
The top-quark spin-correlations have been taken into account in an
approximate way, similar to what has been done in the
\aMCNLO~\cite{Artoisenet:2012st} and \PowHel~\cite{Garzelli:2011vp}
implementations, to allow for studies based on top-quark spin-polarization
effects~\cite{Artoisenet:2012st,Demartin:2014fia,Biswas:2014hwa}.  A
first comparison of this interface with \PowHel{} and \SHERPA{}, that
did not include spin-correlation effects, was presented in the context
of the Les Houches 2013 Workshop~\cite{Butterworth:2014efa}.  After
more detailed cross checks and further improvements, we are now making
our implementation public through the \POWHEGBOX{} website, {\tt
  http://powhegbox.mib.infn.it/}.

In view of the dedicated effort on $t\bar{t}H$ analyses in both the
ATLAS and CMS collaborations, in this paper we would like to present
the details of our implementation and take the opportunity to address
some of the issues that have emerged in recent experimental studies.
In particular, we will investigate the dependence of theoretical
predictions on the choice of a static or dynamical
renormalization/factorization scale, the possibly different behavior
of the interface with \PYTHIA/\HERWIG, and the effect of top-quark
spin-correlations on various kinematic distributions of the decay
products of the $t\bar{t}H$ final state.  Even though the background
to $t\bar{t}H$ represents a major hurdle which needs to be overcome in
order to properly measure this very important production channel, a
very accurate control of the signal is still the first necessary step,
and the possibility for the experimental community to have access to
several public tools is very valuable. Our implementation answers this
need and provides the original NLO-QCD analytic
calculation~\cite{Reina:2001sf,Reina:2001bc,Dawson:2002tg,Dawson:2003zu}
in the same framework (\POWHEGBOX{} in this case) as other processes
that enter the $t\bar{t}H$ studies, from $t\bar{t}$
production~\cite{Frixione:2007nw} to single-top
production~\cite{Alioli:2009je} or Higgs production in gluon-gluon
fusion~\cite{Alioli:2008tz,Bagnaschi:2011tu}, offering a fully
consistent alternative to analogous studies in the
\MadaMCNLO{}~\cite{Alwall:2014hca} framework.

\section{Implementation}
%\subsection{The \POWHEGBOX{} implementation}
\label{sec:powheg-box}
The implementation of a new process in the framework of the
\POWHEGBOX{} requires a list of all independent flavor structures for
the Born and the real-emission contributions, the Born and the
real-emission amplitudes squared, the finite parts of the virtual
amplitudes interfered with the Born amplitude, the spin- and color
correlated amplitudes squared, and a parametrization of the phase
space for the Born process. The tree-level amplitudes can be generated
with the help of the build tool based on {\tt
  MadGraph~4}~\cite{Stelzer:1994ta,Alwall:2007st} that is available in
the \POWHEGBOX{}. For $pp\to \tth$, we use the virtual amplitudes
of~\cite{Reina:2001sf,Reina:2001bc,Dawson:2002tg,Dawson:2003zu} and
adapt them to the format required by the \POWHEGBOX{}.
In order to validate our implementation, we have checked that the
results of \cite{Dawson:2002tg,Dawson:2003zu} are fully reproduced
by the NLO-QCD mode of the \POWHEGBOX{}.

%\subsection{The Implementation of Decays}
%\label{sec:decays}
%
The fixed-order calculation of
\cite{Reina:2001sf,Reina:2001bc,Dawson:2002tg,Dawson:2003zu} provides
NLO-QCD corrections to the production of an on-shell $\tth$ state at a
hadron collider. If implemented in a multi-purpose Monte-Carlo program
like \PYTHIA{}, decays of the scalar Higgs boson can easily be
simulated. In principle, also the decays of the top quarks can be
taken care of externally. However, merely combining the on-shell
calculation for $pp\to\tth$ with a separate, spin-averaged simulation
of the top-quark decays by the shower Monte Carlo results in a loss of
information on correlations between the production and the decay
stages.

A method for overcoming this limitation was proposed in
\cite{Frixione:2007zp} and has by now been applied in several
\POWHEGBOX{} implementations of processes with unstable particles
\cite{Alioli:2009je,Oleari:2011ey,Alioli:2011as}. The basic idea of
this approach is to first produce on-shell $\tth$ events including
NLO-QCD corrections matched to a parton-shower generator (NLO+PS) via
the \POWHEG{} framework, and subsequently simulate the decays
according to a distribution determined by matrix elements for $H (t\to
b\ell^+\nu)(\bar t\to \bar b\ell^-\bar\nu)$ and $H (t\to
b\ell^+\nu)(\bar t\to \bar b\ell^-\bar\nu) + \mr{jet}$ final states
that retain correlations in the top-quark decays, both at leading
order and in the real-emission contributions. In addition to an
improved description of spin correlations between the production and
decay stages, this method allows for a moderate reshuffling of momenta
in such a way that the virtualities of the top quarks are distributed
according to Breit-Wigner shapes. For the actual implementation of
this procedure we follow closely what has been done for the related
case of $t\bar t +\mr{jet}$ production \cite{Alioli:2011as}, and refer
the reader to that reference for further technical details.
The code version we are providing in the \POWHEGBOX{} allows the user
to activate this feature for an improved description of top-quark
decays via a switch in the input file. Alternatively, the code can
generate on-shell events for $pp\to \tth$ that are subsequently
decayed externally, for instance via \PYTHIA{}.

\section{Results}
\label{sec:results}

The results presented in this section have been obtained using a
prototype setup where we consider the LHC operating at
$\sqrt{s}=8$~TeV, with the top quark and Higgs boson masses chosen to
be, respectively, $m_t=172.5$~GeV, and $m_H=125$~GeV. After having
validated the \POWHEGBOX{} implementation against the parton-level
NLO-QCD results of
Refs.~\cite{Reina:2001sf,Reina:2001bc,Dawson:2002tg,Dawson:2003zu}, we
have focused on some of the issues that have been raised in recent
experimental analyses concerning differences encountered in the
behavior of several distributions when interfacing the NLO-QCD
calculation of $pp\rightarrow t\bar{t}H$ with either \PYTHIA{} or
\HERWIG{}. With this respect, results are presented using the {\tt
  PYTHIA-6.4.25}~\cite{Sjostrand:2006za}, {\tt
  PYTHIA-8.183}~\cite{Sjostrand:2007gs}, and {\tt
  HERWIG-6.5.10}~\cite{Corcella:2000bw} releases of the aforementioned
parton-shower algorithms.  We note that, in principle, \POWHEG{}
requires a transverse-momentum ordered parton-shower program, such as
\PYTHIA. In order to correct for unwanted radiation effects that occur
in angular-ordered parton shower programs, such as \HERWIG, a
vetoed-truncated shower has to be included. Since the public version
of \HERWIG{} does not provide a truncated shower, we neglect these
contributions and consider them as part of the systematic
uncertainties of our results.  In all our numerical results,
underlying event and multi-parton interactions are disregarded.  We
note that more specific studies should be developed in the context of
dedicated experimental analyses and we facilitate them by making our
codes public through the \POWHEGBOX{}.

In this section we choose to illustrate the reach of our
implementations by showing a sample of distributions for the main
final-state objects ($t$, $\bar{t}$, and $H$) and their decay
products, obtained by interfacing the NLO-QCD calculation with both
\PYTHIA{} and \HERWIG{}.  Indeed for our numerical analysis, we only
consider the decay products of the top and antitop quarks. Decays of
the scalar Higgs boson (no spin-correlation involved) can easily be
simulated with the external parton-shower Monte Carlo, and we provide
a respective option in the public version of our \POWHEGBOX{}
implementation. 
We study the dependence of
distributions on the choice of either a fixed or a dynamical
renormalization/factorization scale.  We do not take up a study of
other issues, like the dependence on intrinsic parton-shower scales,
or the residual uncertainty due to the choice of Parton Distribution
Functions (PDF). These studies are important and should be performed
in the context of specific experimental analyses. All our results are
presented using the CT10nlo~\cite{Lai:2010vv} set of PDF as implemented
in the LHAPDF library~\cite{Whalley:2005nh} with the number of light
quark flavors set to $N_f=5$ and the strong coupling at NLO QCD
determined by $\alpha_s(M_Z)=0.118$.

To explore the issue of scale dependence, we set the
renormalization and factorization scales equal to each other and
studied both the case of a fixed and a dynamical scale.  When using a
fixed scale we have chosen $\mu_0=m_t+m_H/2$ as central scale, while
we have chosen $\mu_0=(m_T(t)m_{T}(\bar{t})m_T(H))^{1/3}$ as central
value when working with a dynamical scale (where $m_T(P)$ denotes the
transverse mass of particle $P$). The renormalization/factorization
scale dependence of the results is assessed by arbitrarily varying
such scales by factors of $\xi=1/2$ and $2$ about the central value
$\mu_0$. As default choice, we opt for the fixed scale and use
$\mur=\muf = m_t+m_H/2$ unless otherwise specified.

%%%%%%%%%%%%%%%%%%%%%
% NLO vs NLO+PS plots

\begin{figure}[tp]
\begin{center}
\begin{tabular}{lr}
\includegraphics[scale=0.7,height=8truecm,width=8truecm,trim=10 0 70 20,clip]{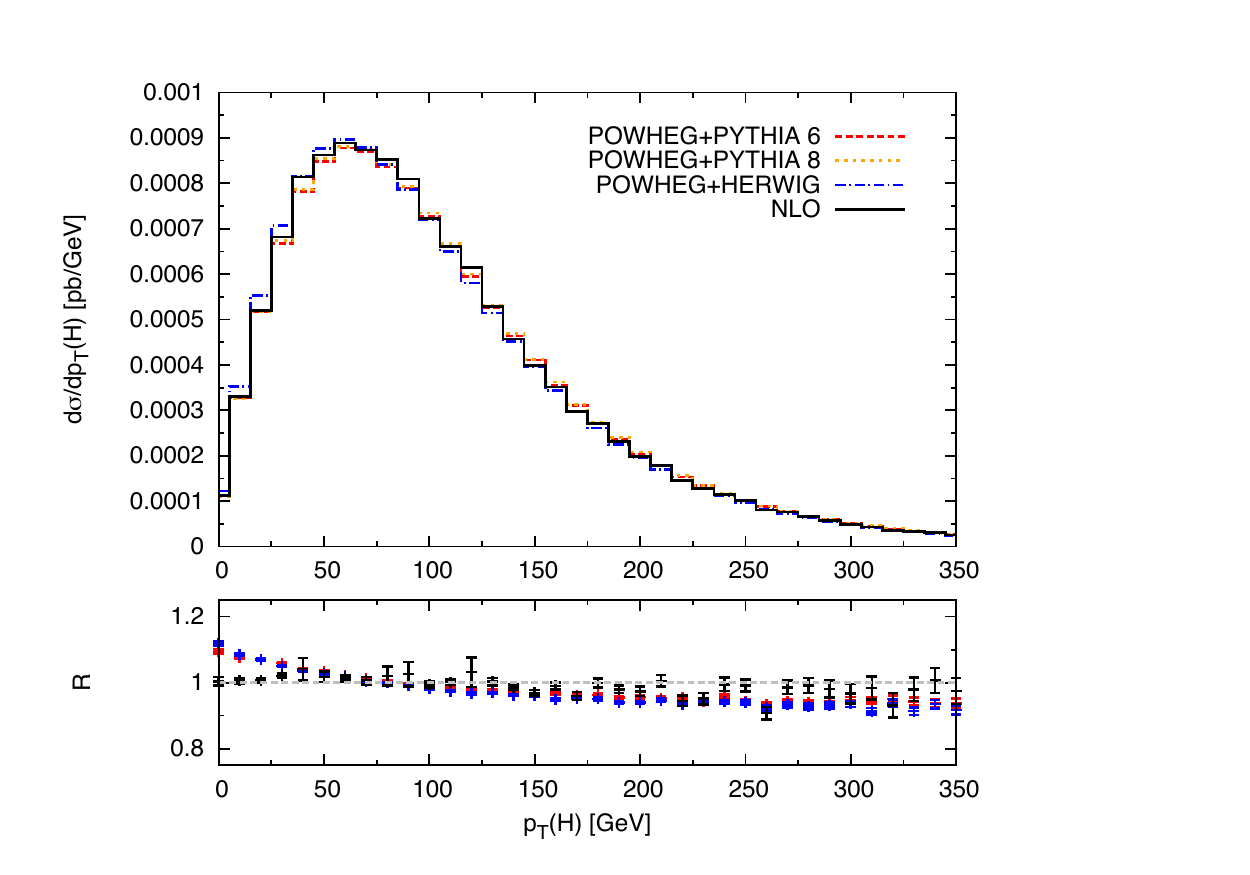} &
\includegraphics[scale=0.7,height=8truecm,width=8truecm,trim=10 0 70 20,clip]{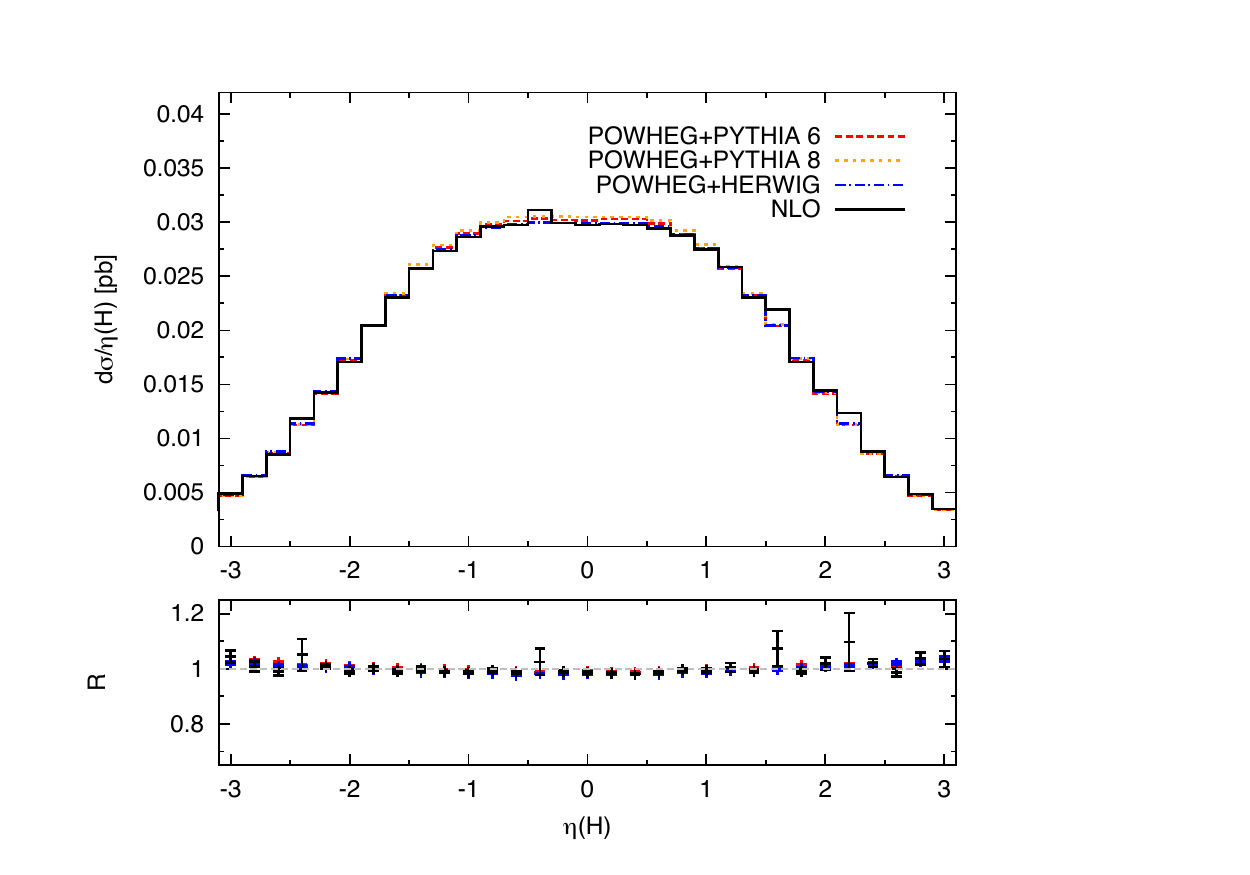}
\end{tabular}
\caption{The $p_T$ (left) and $\eta$ (right) distributions of the
  Higgs boson at NLO-QCD with no parton shower (solid, black), and
  with parton shower as obtained through \POWHEG+\PYTHIAsix{} (long-dashed,
  red), \POWHEG+\PYTHIAeight{} (short-dashed, orange), and \POWHEG+\HERWIG{}
  (dot-dashed, blue) respectively, for a fixed-scale choice (see
  text).  The lower panels show the ratios:
  $R=d\sigma(\mr{NLO})/d\sigma(\PYTHIAsix{})$~(black),
  $R=d\sigma(\HERWIG{})/d\sigma(\PYTHIAsix{})$~(red), and
  $R=d\sigma(\HERWIG{})/d\sigma(\PYTHIAeight{})$~(blue). The error
  bars indicate the statistical uncertainties of the Monte-Carlo
  integration.}
\label{fig:pteta_h_nlo_pythia6-8_herwig}
\end{center}
\end{figure}

\begin{figure}[tp]
\begin{center}
\begin{tabular}{lr}
\includegraphics[scale=0.7,height=8truecm,width=8truecm,trim=10 0 70 20,clip]{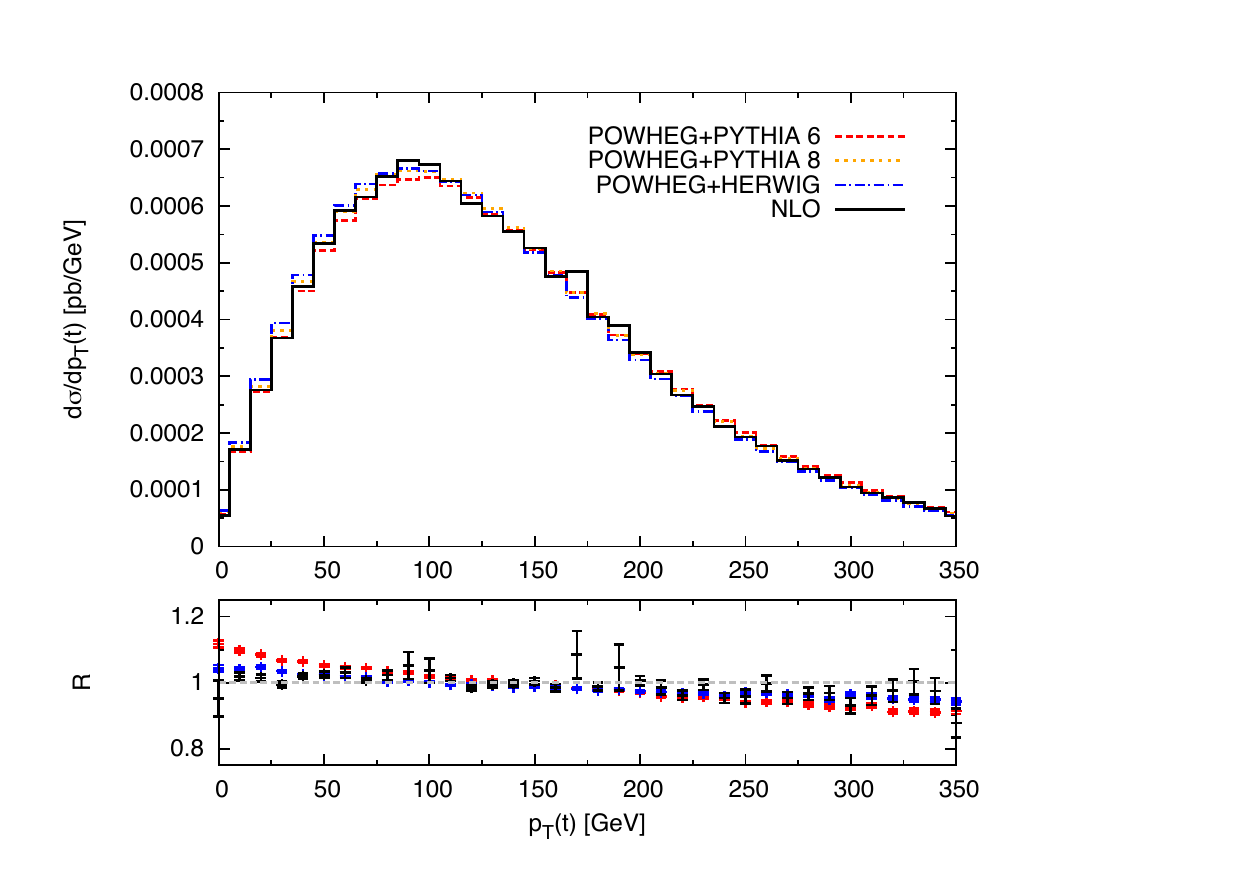}&
\includegraphics[scale=0.7,height=8truecm,width=8truecm,trim=10 0 70 20,clip]{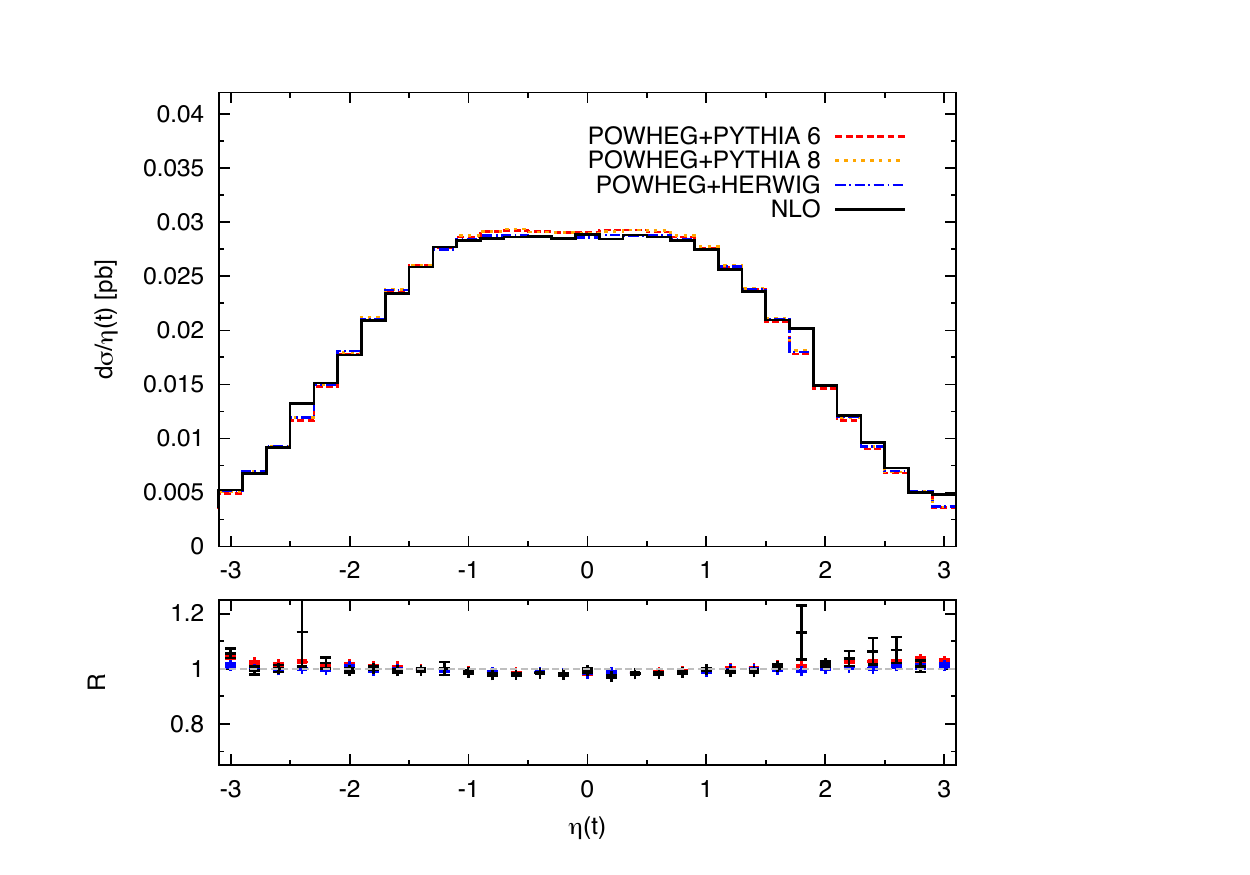}
\end{tabular}
\caption{The $p_T$ (left) and $\eta$ (right) distributions of the top
  quark at NLO-QCD with no parton shower (solid, black), and with
  parton shower as obtained through \POWHEG+\PYTHIAsix{} (long-dashed,
  red), \POWHEG+\PYTHIAeight{} (short-dashed, orange), and \POWHEG+\HERWIG{}
  (dot-dashed, blue) respectively, for a fixed-scale choice (see
  text).  The lower panels show the ratios:
  $R=d\sigma(\mr{NLO})/d\sigma(\PYTHIAsix{})$~(black),
  $R=d\sigma(\HERWIG{})/d\sigma(\PYTHIAsix{})$~(red), and
  $R=d\sigma(\HERWIG{})/d\sigma(\PYTHIAeight{})$~(blue).  The error
  bars indicate the statistical uncertainties of the Monte-Carlo
  integration.}
\label{fig:pteta_t_nlo_pythia6-8_herwig}
\end{center}
\end{figure}

\begin{figure}[tp]
\begin{center}
\begin{tabular}{lr}
\includegraphics[scale=0.7,height=8truecm,width=8truecm,trim=10 0 70 20,clip]{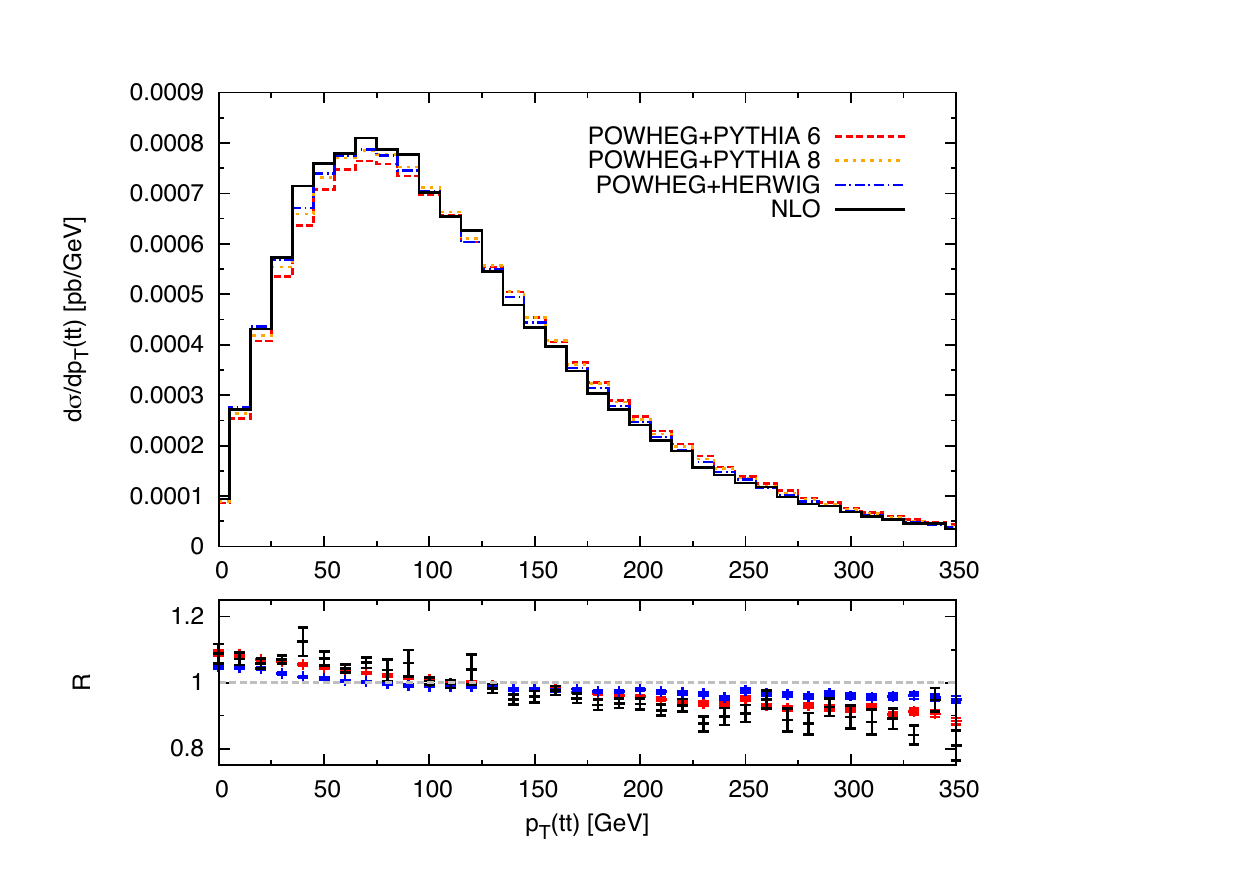}&
\includegraphics[scale=0.7,height=8truecm,width=8truecm,trim=10 0 70 20,clip]{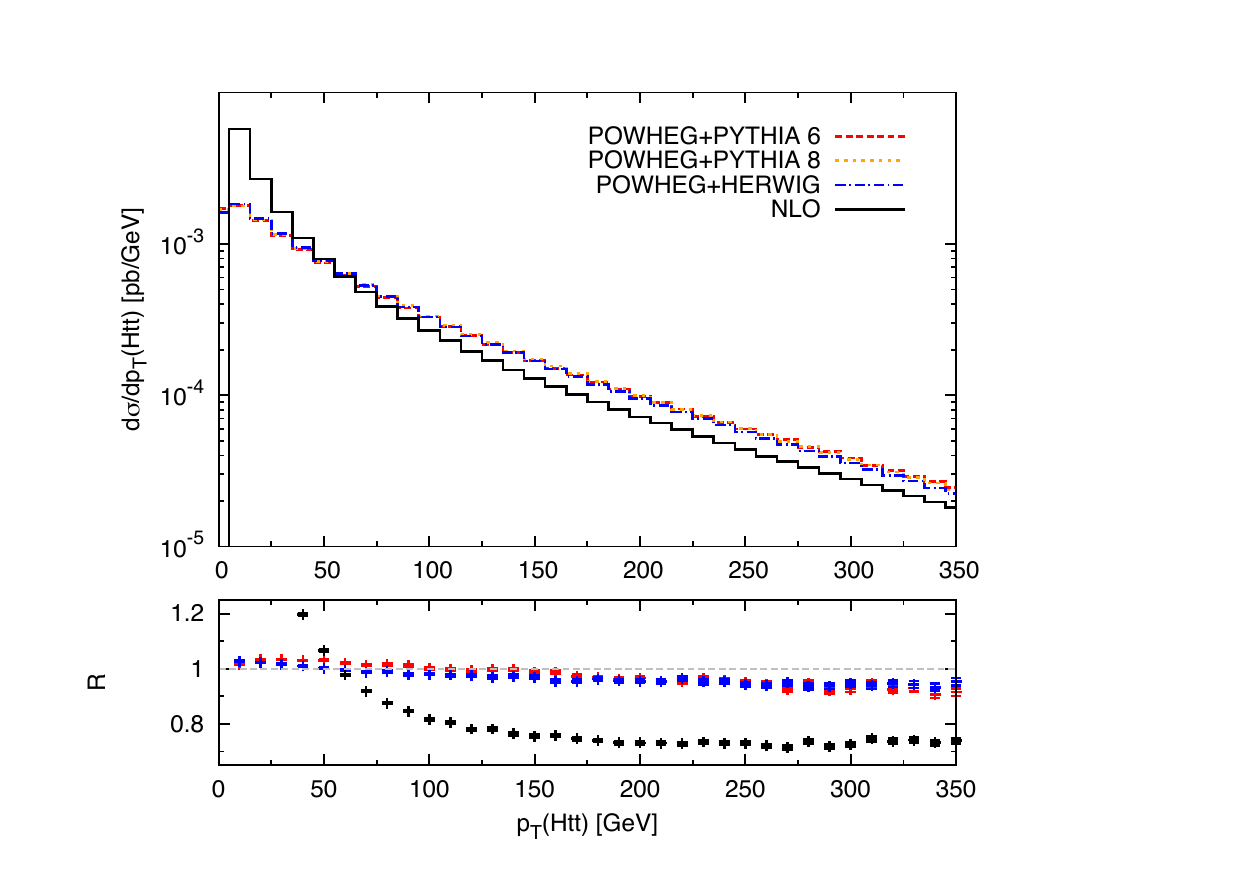}
\end{tabular}
\caption{The $p_T$ distributions of the $t\bar t$ pair (left) and of
  the $t\bar t H$ system (right) at NLO-QCD with no parton shower
  (solid, black), and with parton shower as obtained through
  \POWHEG+\PYTHIAsix{} (long-dashed, red), \POWHEG+\PYTHIAeight{} (short-dashed,
  orange), and \POWHEG+\HERWIG{} (dot-dashed, blue) respectively, for a
  fixed-scale choice (see text).  The lower panels show the ratios:
  $R=d\sigma(\mr{NLO})/d\sigma(\PYTHIAsix{})$~(black),
  $R=d\sigma(\HERWIG{})/d\sigma(\PYTHIAsix{})$~(red), and
  $R=d\sigma(\HERWIG{})/d\sigma(\PYTHIAeight{})$~(blue).  The error
  bars indicate the statistical uncertainties of the Monte-Carlo
  integration.}
\label{fig:pttth_nlo_pythia6-8_herwig}
\end{center}
\end{figure}

To study the impact of different parton-shower
  generators on the NLO-QCD parton-level results, we present in
  Figs.~\ref{fig:pteta_h_nlo_pythia6-8_herwig}-\ref{fig:pttth_nlo_pythia6-8_herwig}
  the comparison between the parton-level NLO-QCD results and the
  corresponding results obtained upon matching with either
  \PYTHIAsix, \PYTHIAeight, or \HERWIG{} as implemented in the
  \POWHEGBOX{} framework.
In
Figs.~\ref{fig:pteta_h_nlo_pythia6-8_herwig}-\ref{fig:pteta_t_nlo_pythia6-8_herwig}
we illustrate the impact of the parton shower on the fixed-order
NLO-QCD results for the transverse-momentum and rapidity distributions
of the Higgs boson and the top quark, respectively. Parton-shower
effects do not cause large changes compared to the NLO results of
these distributions, and differences between the \PYTHIA{} and the
\HERWIG{} implementations are small. We noticed however a systematic
enhancement of the low $p_T$ region in \HERWIG{} with respect to
\PYTHIAsix{}, while \PYTHIAeight{} seems to have a better agreement
with \HERWIG{} over the entire $p_T$ spectrum.
The right panel of Fig.~\ref{fig:pttth_nlo_pythia6-8_herwig} shows the
transverse momentum of the $\tth$ system. This observable is entirely
due to real-radiation contributions in the NLO-QCD corrections and
parton-shower effects. At leading order, the transverse momentum of
the $\tth$ system is zero, because of momentum conservation. If
real-emission contributions are taken into account, as in the
fixed-order NLO-QCD calculation, this observable exhibits a divergence
as the transverse momentum of the entire system approaches zero.
In the NLO+PS result, this
behavior is tamed by a Sudakov factor. The \POWHEG+\PYTHIA{} and
\POWHEG+\HERWIG{} results are thus much better behaved in the region
of low $p_T(\tth)$ that correspond to the emission of a soft jet, and
they are compatible over the full $p_T$ spectrum.  We
  could present distributions for jet observables as well, starting
  from the $p_T$ and $\eta$ of the hardest or next-to-hardest jets,
  but they would not help us to judge the impact of the parton shower
  if we did not set in place, at the same time, more specific
  selection cuts aimed at distinguishing the first emission described
  by the hard matrix elements or the top-quark decays from the
  following emissions coming from the parton shower, as well as cuts
  and vetoes aimed at distinguishing light jets from heavy-flavor
  jets. Since these sort of requirements only make sense in the
  context of dedicated experimental analyses, we refrain from
  considering specific jet variables and present only results for
  observables of systems, like the $t\bar{t}H$ system, that carry a
  clear imprint of the jet from the first QCD emission, and allow us
  to verify expected effects like the one due to the Sudakov factor.

\begin{figure}[tp]
\begin{center}
\begin{tabular}{lr}
\includegraphics[scale=0.7,height=8truecm,width=8truecm,trim=10 0 70 20,clip]{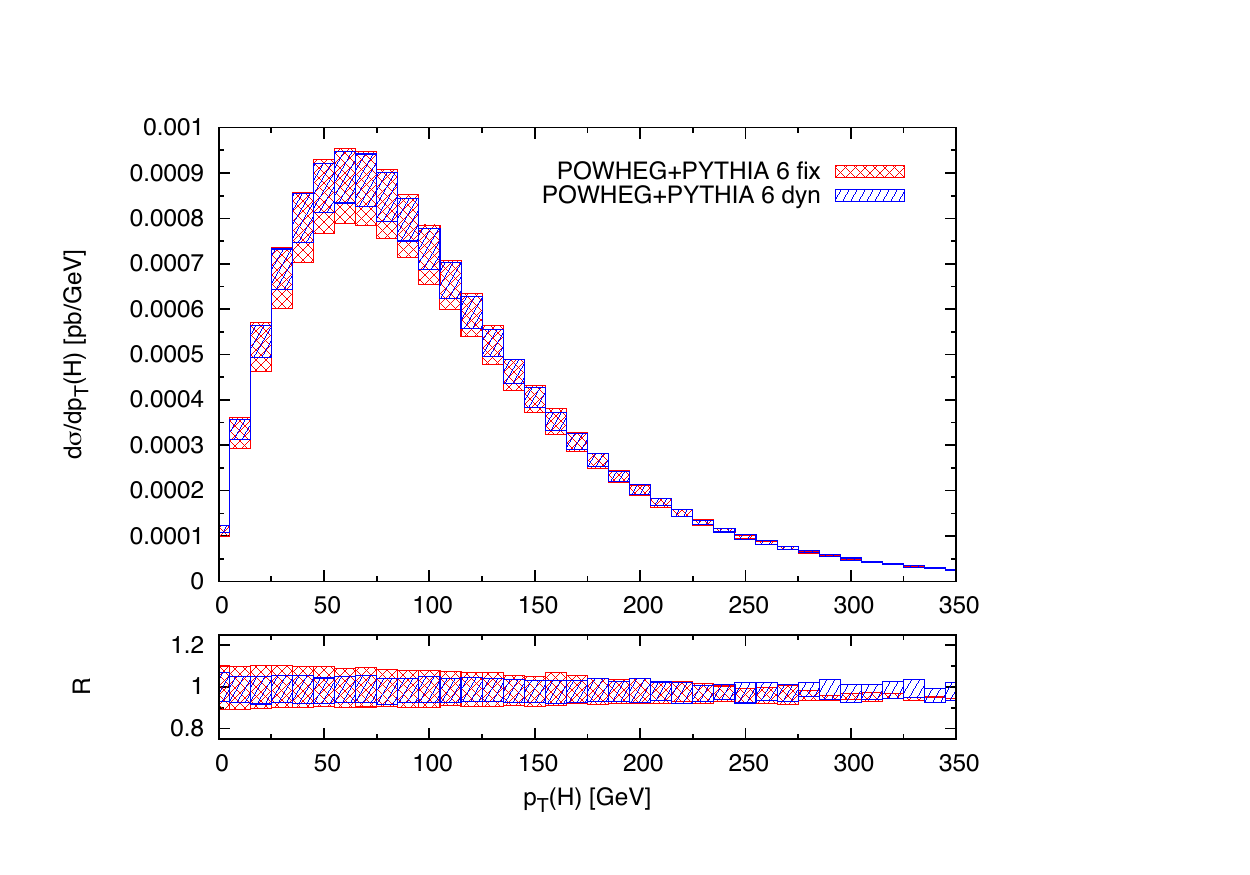}&
\includegraphics[scale=0.7,height=8truecm,width=8truecm,trim=10 0 70 20,clip]{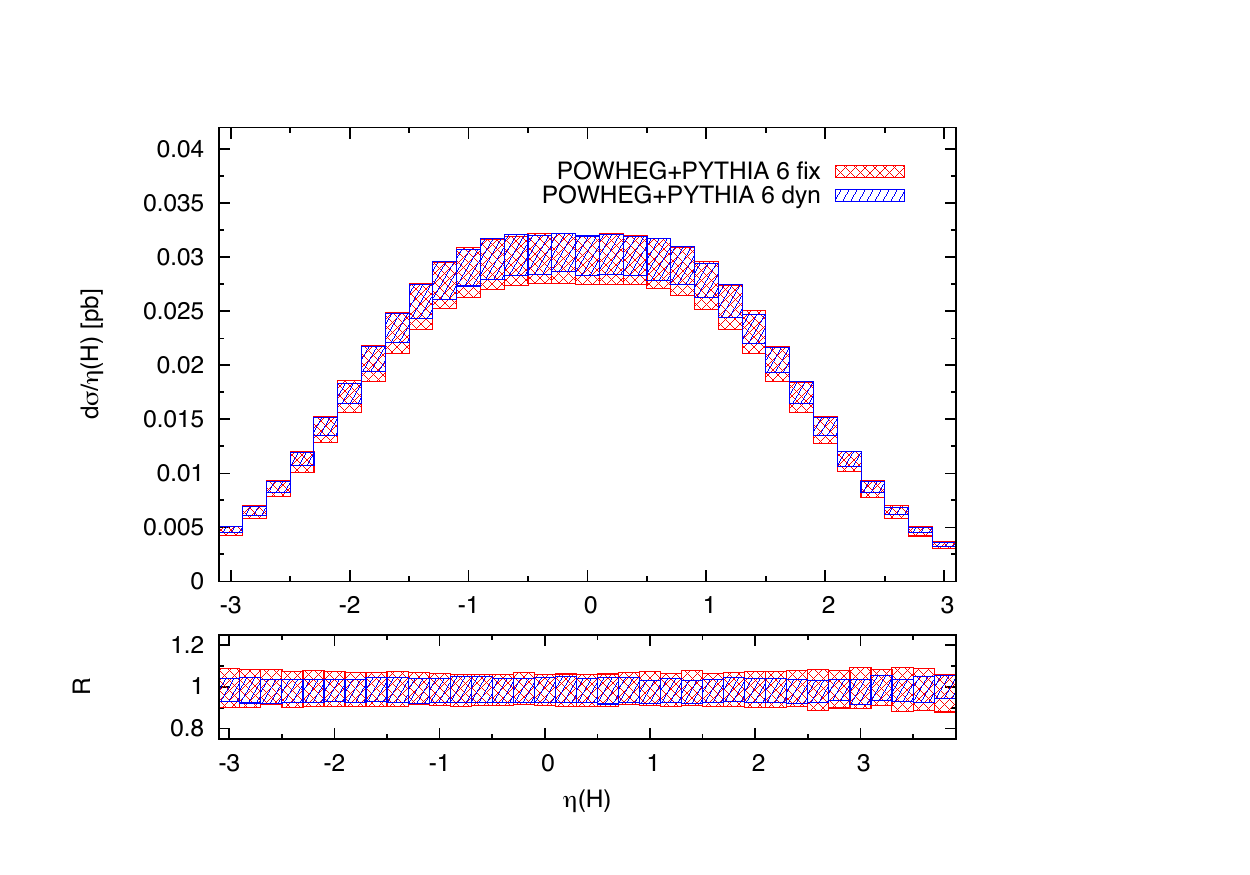}
\end{tabular}
\caption{The $p_T$ (left) and $\eta$ (right) distributions of the
  Higgs boson obtained with \POWHEG+\PYTHIAsix{} for fixed ($fix$) and
  dynamical ($dyn$) renormalization/factorization scales. The lower
  panels show the respective ratios
  $R=d\sigma(\xi\mu_0)/d\sigma(\mu_0)$ for $\xi=(0.5;2)$.}
\label{fig:pteta_h_scale_dep}
\end{center}
\end{figure}

\begin{figure}[tp]
\begin{center}
\begin{tabular}{lr}
\includegraphics[scale=0.7,height=8truecm,width=8truecm,trim=10 0 70 20,clip]{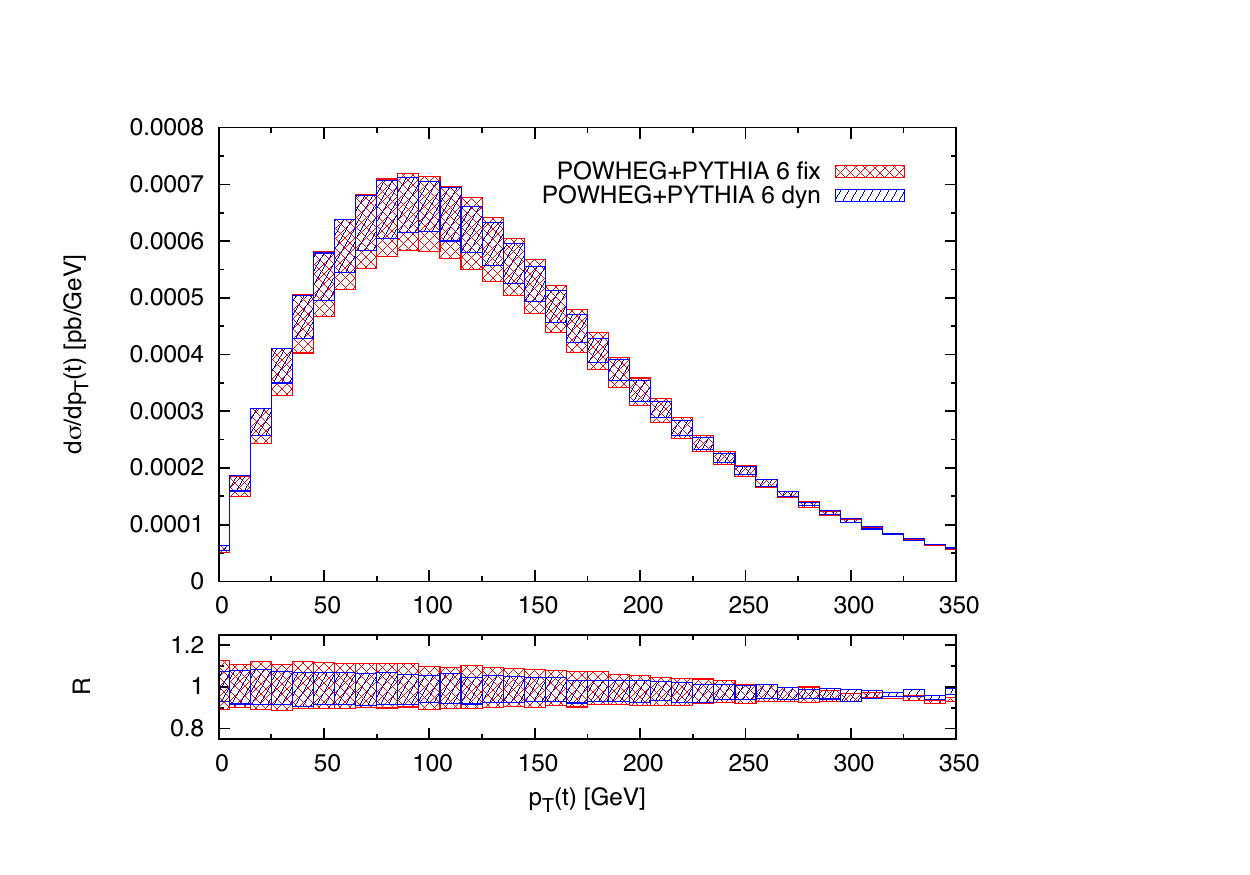}&
\includegraphics[scale=0.7,height=8truecm,width=8truecm,trim=10 0 70 20,clip]{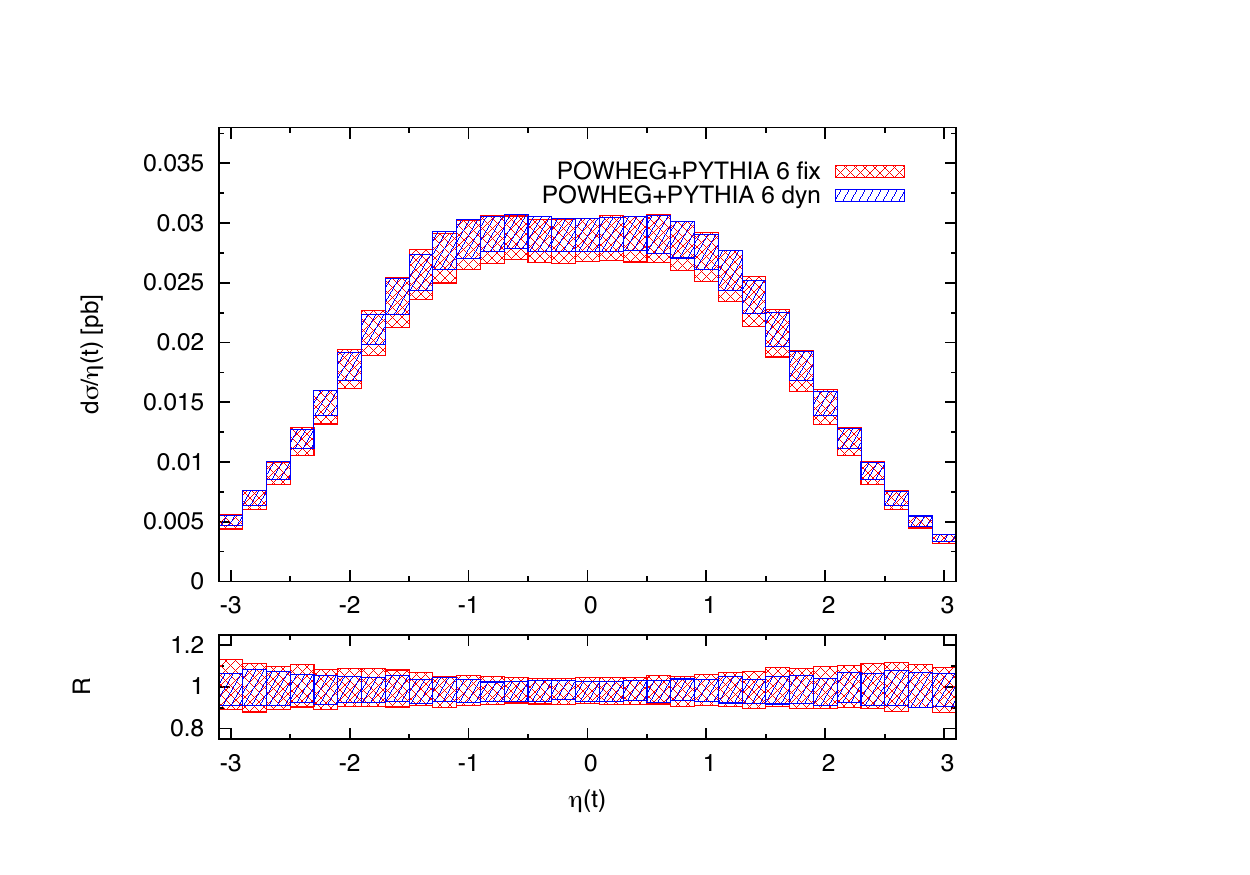}
\end{tabular}
\caption{The $p_T$ (left) and $\eta$ (right) distributions of the top
  quark obtained with \POWHEG+\PYTHIAsix{} for fixed ($fix$) and
  dynamical ($dyn$) renormalization/factorization scales. The lower
  panels show the respective ratios $R=d\sigma(\xi
  \mu_0)/d\sigma(\mu_0)$ for $\xi=(0.5;2)$.}
\label{fig:pteta_t_scale_dep}
\end{center}
\end{figure}

\begin{figure}[tp]
\begin{center}
\begin{tabular}{lr}
\includegraphics[scale=0.7,height=8truecm,width=8truecm,trim=10 0 70 20,clip]{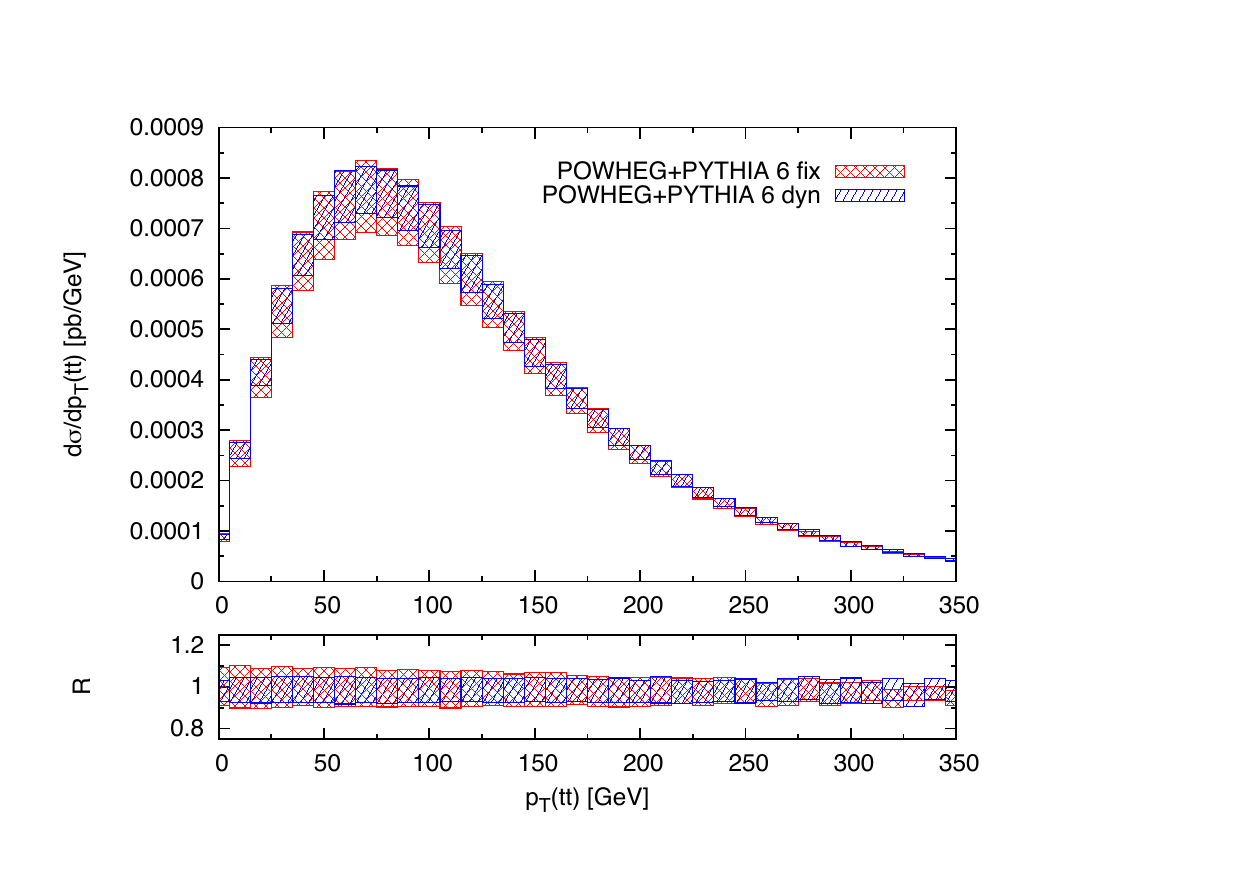}&
\includegraphics[scale=0.7,height=8truecm,width=8truecm,trim=10 0 70 20,clip]{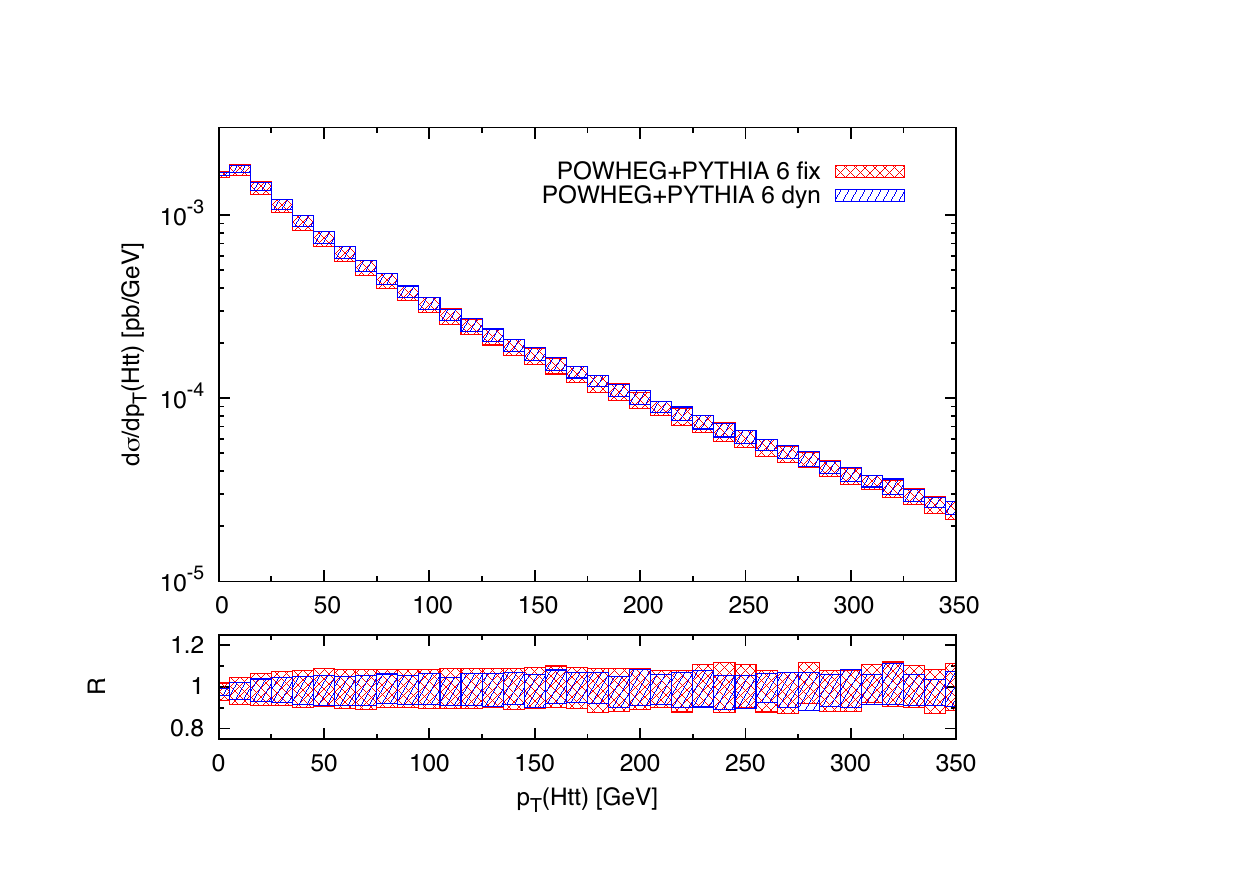}
\end{tabular}
\caption{The $p_T$ distributions of the $t\bar t$ pair (left) and of
  the $t\bar t H$ system (right) obtained with \POWHEG+\PYTHIAsix{}
  for fixed ($fix$) and dynamic ($dyn$) renormalization/factorization
  scales. The lower panels show the respective ratios $R=d\sigma(\xi
  \mu_0)/d\sigma(\mu_0)$ for $\xi=(0.5;2)$.}
\label{fig:pttth_scale_dep}
\end{center}
\end{figure}

%%%%%%
% scale dependence plots:
%
In order to assess the theoretical uncertainties associated with the
choice of renormalization and factorization scale, we have computed
the previously considered distributions for different choices of 
scale as previously explained. In particular,
Figs.~\ref{fig:pteta_h_scale_dep}-\ref{fig:pttth_scale_dep} are
obtained using our \POWHEG+\PYTHIAsix{} implementation.  More
specifically, Figs.~\ref{fig:pteta_h_scale_dep} and
\ref{fig:pteta_t_scale_dep} show the transverse momentum and rapidity
distributions of the Higgs boson and the top quark, respectively. The
scale dependence of the results is considerable, amounting to more
than $\pm 10\%$ in some regions of phase space. Using a dynamical
rather than a fixed scale helps in slightly reducing the scale
uncertainty of the NLO+PS results in all observables.

A similar behavior can be observed in the transverse momentum
distributions of the $t\bar t$~pair and the $\tth$~system,
c.f.~Fig.~\ref{fig:pttth_scale_dep}.
Since the latter observable vanishes at leading order, we expect it to
be plagued by larger uncertainties than distributions that are
genuinely described at NLO accuracy. 

\begin{figure}[tp]
\begin{center}
\begin{tabular}{lr}
\includegraphics[scale=0.7,height=8truecm,width=8truecm,trim=10 0 70 20,clip]{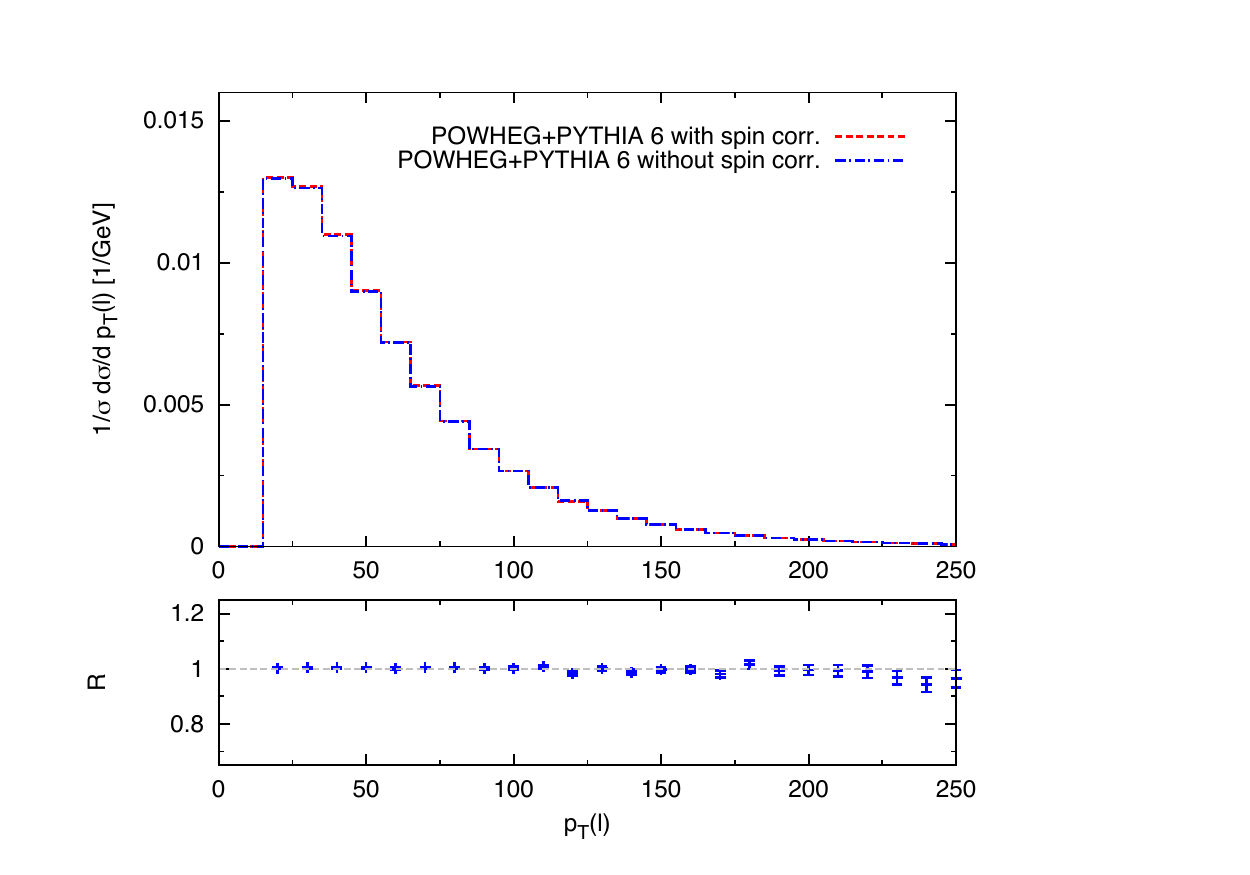}&
\includegraphics[scale=0.7,height=8truecm,width=8truecm,trim=10 0 70 20,clip]{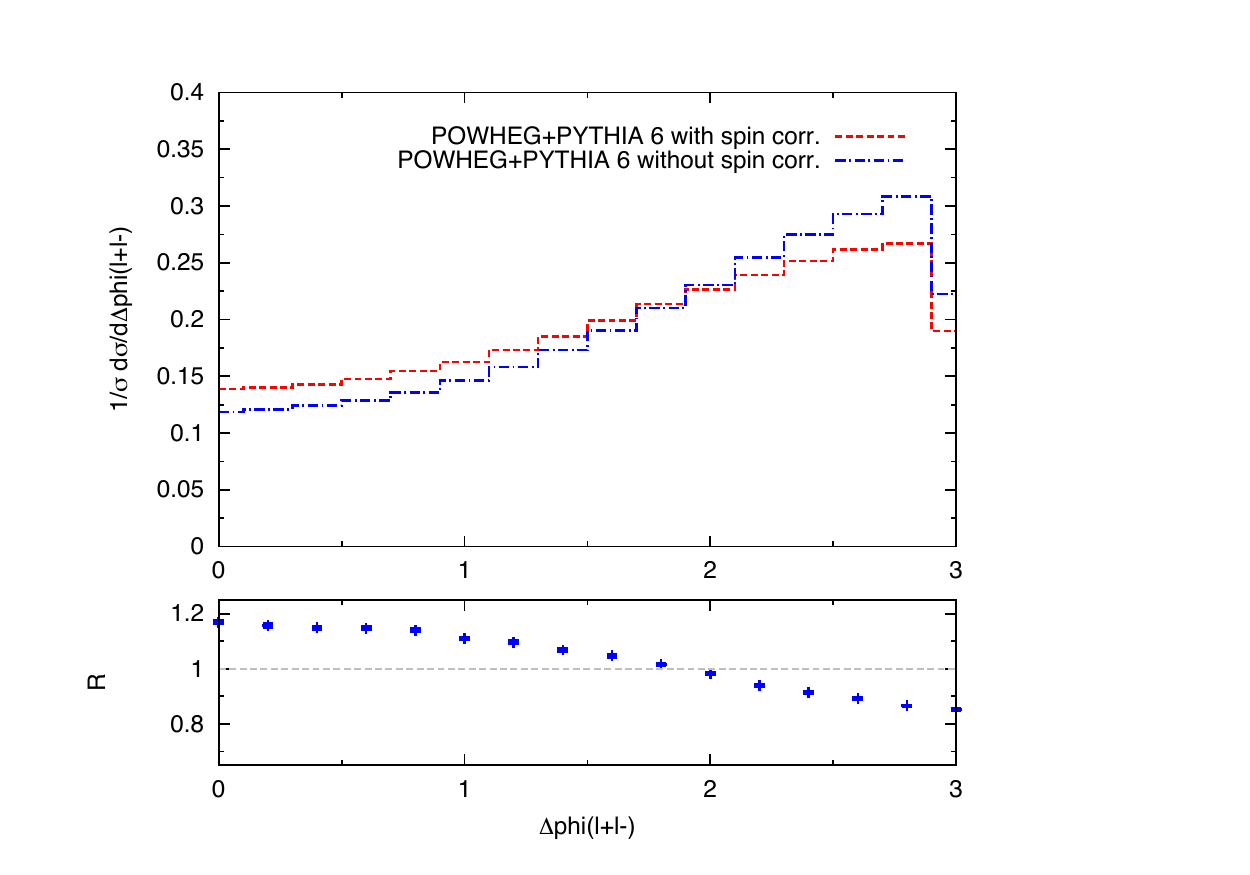}
\end{tabular}
\caption{Normalized transverse-momentum distribution of the hardest
  lepton (left) and azimuthal angle separation of the two hardest
  oppositely charged leptons (right) obtained with
  \POWHEG+\PYTHIAsix{} with (blue lines) and without (red lines) spin
  correlations in the decays of the top quarks, for a dynamic-scale
  choice (see text). The lower panels show the ratios of the results
  without and with spin correlations.}
\label{fig:ptphil}
\end{center}
\end{figure}

%%%%%%
%
% decay correlations:
%
We finally illustrate the impact of spin-correlation effects.  As
pointed out, for instance in Ref.~\cite{Biswas:2014hwa}, exploiting
polarization effects in the top-quark decays can appreciably improve
the sensitivity of the LHC in the $\tth$ channel.  At the same time,
spin-correlation observables can probe the top-Higgs coupling in
$\tth$ production at the LHC, as studied for instance in
Ref.~\cite{Boudjema:2015nda}.  To allow for the simulation of spin
correlations in the decays of the top quarks our \POWHEGBOX{}
implementation resorts to the prescription of
Ref.~\cite{Frixione:2007zp}, as explained in
Sec.~\ref{sec:powheg-box}. NLO-QCD corrections are considered for the
$\tth$ production process only. Improving on the accuracy of the decay
process would require more advanced techniques, such as the method
that has recently been presented for top quark pair production at the
LHC~\cite{Campbell:2014kua}. For processes with additional particles
in the final state, such as $\tth$ production, such a procedure has
never been applied so far and we restrict ourselves to the more
approximated description of the decay process for the time being.
To explore the impact of spin-correlation effects on experimentally
accessible observables, we consider the leptonic decay modes of the
top quarks and focus on final states with at least two oppositely
charged leptons with
\begin{equation}
p_{T,\ell} > 20~\mr{GeV}\,,\quad
|y_\ell| < 2.5\,,
\end{equation}
that are well-separated by $\Delta R_{\ell,j}>0.4$ from identified
jets with $p_{T,j}>20$~GeV and $|y_j|<4.5$.
As illustrated by Fig.~\ref{fig:ptphil}, 
the transverse-momentum distribution of a hard lepton barely depends
on the treatment of the top-quark decays. Predictions for angular
correlations between the two hardest oppositely-charged leptons,
however, exhibit a sizable modification when the simulation of the
top-quark decays is improved.

\section{Conclusions}
\label{sec:conclusions}
In this work we have presented a new, public implementation of $\tth$
production at the LHC at NLO-QCD matched to parton showers in the
framework of the \POWHEGBOX{}. In order to illustrate the reach of the
implementation, we discussed several sources of theoretical
uncertainties, in particular scale dependencies and differences due to
the specific shower Monte Carlo program used.  We found that, within
the limits of such a general study, for most observables differences
between predictions obtained with \PYTHIAsix, with \PYTHIAeight, and
with \HERWIG{} are small. We remind the reader that the public version
of \HERWIG{} that we have used does not have any vetoed truncated
showers and therefore the comparison between \PYTHIA{} and \HERWIG{}
will have to be further investigated in particular in analyses that
include jet observables.  Scale uncertainties are generally
non-negligible and can amount to more than $\pm 10\%$ in some regions
of phase space. The uncertainties obtained using a dynamical and a
fixed scale are compatible, although typically they are smaller when
using a dynamical scale. For observables involving the decay products
of the top quark, we recommend using a code version that does take
into account spin correlations between the production and decay
stages.

\section*{Acknowledgements}
We are grateful to Carlo Oleari for his assistance in making this code
publicly available on the \POWHEGBOX{} website.  This work was
supported in part by the National Science Foundation under Grant
No. PHYS-1066293 and the Aspen Center for Physics.  The work of
B.~J.~is supported in part by the Institutional Strategy of the
University of T\"ubingen (DFG, ZUK~63).  The work of L.~R.~is
supported in part by the U.S. Department of Energy under grant
DE-FG02-13ER41942.  The work of D.~W. is supported in part by the
U.S. National Science Foundation under award no.~PHY-1118138.

\bibliography{tth}
\end{document}